\documentclass[12pt]{article}
\usepackage{amsfonts,epsfig,latexsym,amscd}

\newcommand{\R}{{\mathbb{R}}}

\newcommand{\I}{{\mathbb{I}}}

\newcommand{\beq}{\begin{equation}}
\newcommand{\eeq}{\end{equation}}
\newcommand{\bea}{\begin{eqnarray}}
\newcommand{\eea}{\end{eqnarray}}

\newcommand{\cd}{\partial}
\newcommand{\wt}{\widetilde}

\newcommand{\betvec}{\mbox{\boldmath{$\beta$}}}
\newcommand{\xivec}{\mbox{\boldmath{$\xi$}}}
\newcommand{\alvec}{\mbox{\boldmath{$\alpha$}}}

\newcommand{\zerovec}{\mbox{\boldmath{$0$}}}
\newcommand{\Li}{L_{\rm int}}
\newcommand{\xv}{{\bf x}}
\newcommand{\yv}{{\bf y}}
\newcommand{\zv}{{\bf z}}
\newcommand{\uv}{{\bf u}}
\newcommand{\jv}{{\bf j}}
\newcommand{\kv}{{\bf k}}
\newcommand{\nv}{{\bf n}}
\newcommand{\rv}{{\bf r}}
\newcommand{\Rv}{{\bf R}}
\newcommand{\Av}{{\bf A}}
\newcommand{\yd}{\dot{{\bf y}}}
\newcommand{\zd}{\dot{{\bf z}}}
\newcommand{\ydd}{\ddot{{\bf y}}}
\newcommand{\zdd}{\ddot{{\bf z}}}

\newcommand{\tr}{{\rm tr}}
\newcommand{\prl}{{\|}}
\newcommand{\gmet}{g}
\newcommand{\ric}{R}
\newcommand{\scal}{Scal}

\newcommand{\be}{\begin{equation}}
\newcommand{\ee}{\end{equation}}

\def\M {{\cal M}}
\def\bx {\bf x}
\def\pr {\partial}
\def\half {\frac{1}{2}}

\begin{document}
\begin{titlepage}
\title{
{\bf Asymptotic Interactions of Critically Coupled Vortices}
}
\vspace{1cm}
\author{{N. S. Manton}\thanks{e-mail address: N.S.Manton@damtp.cam.ac.uk}\\
{\sl Department of Applied Mathematics and Theoretical Physics}\\
{\sl University of Cambridge} \\
{\sl Wilberforce Road, Cambridge CB3 0WA, England}\\
and\\
{J. M. Speight}\thanks{e-mail address: J.M.Speight@leeds.ac.uk}\\
{\sl Department of Pure Mathematics}\\
{\sl University of Leeds} \\
{\sl Leeds LS2 9JT, England} \\ \\
Short title:
Asymptotic Interactions of Vortices}
\date{May, 2002}
\maketitle
\thispagestyle{empty}
\vspace{1cm}
\newpage
\begin{abstract}
\noindent
At critical coupling, the interactions of Ginzburg-Landau vortices
are determined by the metric on the moduli space of static
solutions. Here, a formula for the asymptotic metric for
two well separated vortices is obtained, which depends on
a modified Bessel function. A straightforward extension gives the metric for
$N$ vortices.  The asymptotic metric is also shown to
follow from a physical model, where each vortex is
treated as a point-like particle carrying a scalar charge and a
magnetic dipole moment of the same magnitude.
The geodesic motion of two well separated vortices is investigated, and
the asymptotic dependence of the scattering angle on the impact
parameter is determined. Formulae for the asymptotic Ricci and scalar
curvatures of the $N$-vortex moduli space are also 
obtained.
\end{abstract}
\end{titlepage}

\section{Introduction}
\label{sec:intro}

Critically coupled Ginzburg-Landau vortices and BPS monopoles are two
of the most studied examples of topological solitons in field theory
\cite{JT}. Vortices are particle-like solutions of the abelian Higgs 
theory in
two dimensions and BPS monopoles are solutions of a Yang-Mills-Higgs
theory in three dimensions.

At critical coupling (separating the Type
I and Type II regimes of superconductivity), vortices exert no static
forces on each other, and there are static multi-vortex
solutions. These satisfy the planar Bogomolny equations \cite{Bo}
\begin{eqnarray}
D_1\phi + i D_2\phi &=& 0 \label{Bog1}\\
B + \half (|\phi|^2 -1) &=& 0 \label{Bog2} \,,
\end{eqnarray}
and the boundary condition $|\phi| \to 1$
as $|{\bf x}| \to \infty$. Here, $D_i \phi = \pr_i \phi + i A_i \phi$ is
the covariant derivative of the
complex scalar field $\phi$, and $B = \pr_1 A_2 - \pr_2 A_1$ is the
magnetic field in the plane.
Taubes showed that an $N$-vortex solution is uniquely determined by
specifying $N$ points where $\phi$ is zero \cite{Tau1,JT}. The moduli
space of $N$-vortex solutions, $\M_N$, is therefore
the configuration space of $N$ unordered points in the plane, which is a smooth
$2N$-dimensional manifold. Using the first Bogomolny equation, the
gauge potential can be eliminated, and the second equation can then be
written in terms of the gauge invariant field $h = \log|\phi|^2$ as
\be
\nabla^2 h - e^h + 1 = 4\pi \sum^N_{r=1} \delta (\bx - {\bf y}_r) \,.
\label{heqn1}
\ee
The points $\{ {\bf y}_r  : 1 \le r \le N \}$ are the
locations of the vortices, where $\phi$ vanishes and $h$ has a
logarithmic singularity. The boundary condition is $h \to 0$
as $|{\bf x}| \to \infty$. Taubes showed that $h$ approaches 0 
exponentially fast \cite{JT}.

Static BPS monopoles are solutions in $\R^3$ of the Bogomolny
equations
\be
B_i = D_i \Phi \,,
\label{MonoBog}
\ee
where $D_i \Phi$ is the covariant derivative
of the adjoint Higgs field and $B_i$
is the Yang-Mills magnetic field. For fields of finite energy there is
a well-defined monopole number $N$, and the moduli space of $N$-monopole
solutions for gauge group $SU(2)$ is a $4N$-dimensional smooth
manifold. It is not so simple
as in the vortex case to say precisely what the moduli signify without
introducing some additional structures (e.g. Donaldson's rational
map), but for well separated monopoles there are four moduli
associated with each of them. Three specify the location of the
monopole, and the fourth is an internal phase angle.

We are interested not just in static solutions but also in
time-dependent ones. We suppose the complete Lagrangian, both for
vortices and for monopoles, is the Lorentz invariant extension of the
static energy function, with a kinetic term quadratic in the time
derivatives of the fields. In the vortex case, the Lagrangian density is that
of the abelian Higgs theory at critical coupling
\be
{\cal L} = -\frac{1}{4} F_{\mu\nu}F^{\mu\nu}+\half \overline{D_\mu \phi}
D^\mu \phi - \frac{1}{8} (|\phi|^2 -1)^2 \,,
\label{abHig}
\ee
where $F_{\mu\nu} = \pr_\mu A_\nu - \pr_\nu A_\mu$ and $D_\mu \phi
= \pr_\mu \phi + i A_\mu \phi$. The vortices can move at arbitrary
speeds less than the speed of light. For monopoles the situation is
similar, but the theory is non-abelian. High speed collisions of
either vortices or monopoles
are complicated, involving substantial energy transfer to
radiation modes of the fields, and amenable only to numerical
simulation. However, collisions at slow speeds can be treated
adiabatically, using the geodesic approximation \cite{Ma1}. The idea here
is that the moduli space of static solutions acquires a natural
metric by restricting the kinetic terms of the field theory
Lagrangian to motion tangent to the moduli space, and it can
be shown that the geodesic trajectories on moduli space accurately
model the field theory dynamics of the solitons. An argument for
this was given in \cite{Ma1}. It was justified rigorously for vortex
motion \cite{Stu1} and for monopole motion \cite{Stu2} by Stuart. It is
fairly clear now that the geodesic approximation is the formal
non-relativistic limit of the field theory dynamics of the solitons,
where radiation is neglected.

Having recognized the importance of the metric on moduli space, it
becomes desirable to calculate it. This is not so easy. For
monopoles there is an explicit understanding of the metric only for
one monopole, where the metric is flat, and for two monopoles, where
it was calculated by Atiyah and Hitchin \cite{AH}. For vortices there is a
general formula due to Samols \cite{Sa}, which we shall use below.
Again, for one vortex the metric is flat, but even for two vortices the
metric is not known explicitly.

However, it is possible to calculate the explicit asymptotic
metric on the moduli space for $N$ well separated monopoles. There are now
two approaches to this calculation. The first is physically motivated,
and not quite rigorous \cite{Ma2, GM2}. The monopoles are treated as
point-like objects, carrying a magnetic charge and a scalar charge of 
equal magnitude. These charges are regarded as sources for auxiliary
linear fields through which the monopoles interact.
(Monopoles, in reality, are smooth but nonlinear, with a core radius
of order 1, but provided their separation is much greater than 1, a
linearization of the fields appears justified.) For monopoles 
at rest, the magnetic forces exactly cancel the scalar forces, so
there is no net force. For monopoles in relative
motion, the magnetic and scalar forces are not identical, because of
their different Lorentz transformation properties, and there are net forces
which cause the monopoles to scatter non-trivially. The effects can
be encapsulated in an $N$-particle Lagrangian with a purely kinetic
term, quadratic in velocities. The coefficient matrix of this quadratic
form defines the metric on the moduli space for well separated
monopoles. An alternative approach is due to Bielawski, who calculated
rigorously the asymptotic form of the Nahm data associated with well
separated monopoles, and from this calculated the asymptotic metric on moduli
space \cite{Bi}. These approaches give the same result, and they are
consistent with the
Atiyah-Hitchin metric for two monopoles, whose asymptotic form was derived
in \cite{GM1}. The asymptotic $N$-monopole metric, like the true
metric on the $N$-monopole moduli space, is hyperk\"ahler, but unlike
the true metric it has singularities when the monopoles come close together.

These results on monopoles motivated the present work. Here we give an
explicit expression for the metric on the
$N$-vortex moduli space, for $N$ well separated vortices.
Furthermore we calculate it in two ways. Our first approach is to
take Samols' general formula and evaluate
the quantities occurring there by a method of matched asymptotic
expansions. Essentially, we solve eq.\ (\ref{heqn1}) for two well separated
vortices, calculating the effect of one vortex on the other at
linear order, and from the solution determine the asymptotic 2-vortex
metric. It is straightforward to generalize the 2-vortex metric
to the $N$-vortex metric. 
We need to assume that for well separated vortices the field
$h$ far from the vortex cores obeys the linearization of
(\ref{heqn1}), namely the Helmholtz equation
\be
\nabla^2 h - h = 0 \,,
\label{helm0}
\ee
and that the relevant solution is a linear superposition of the
solutions due to the $N$ vortices separately. Corrections due to
the nonlinear terms neglected in eq. (\ref{helm0}) are of higher
exponential order in the separations. However, a careful treatment of
this point is lacking, and would require a considerable refinement of 
Taubes' estimate of the exponential decay of solutions. 

Our second approach is the more physical. It is a variant of the
calculation involving point-like monopoles, and regards well separated
vortices as point-like sources interacting via auxiliary linear fields. A
study of the static forces between vortices that are close to critical
coupling shows that well separated vortices can be regarded as
particles each carrying
a scalar charge and a magnetic dipole moment (thought of as perpendicular to
the plane which the vortex inhabits) \cite{Sp}. For critically
coupled vortices, the magnitudes of the scalar charge and the dipole
are the same, and the static forces due to them cancel. For vortices
in motion, the scalar and magnetic forces do not exactly cancel, but
result in velocity-dependent forces. Again there is an effective
Lagrangian for two well separated vortices which is purely kinetic,
and from this the metric can be read off. The extension to
$N$ vortices is as before.

The asymptotic metric for vortices, which involves the Bessel function
$K_0$, has some similarities
to the true metric. It has the same isometries, and like the true
metric, it is K\"ahler. As for monopoles, the asymptotic
metric becomes singular as the vortices approach one another closely,
since it is not positive definite if the minimum vortex separation is below
a certain critical value ($2.21$ to two decimal places in the case $N=2$). 
Of course, the asymptotic metric is not valid in this region.

It remains open to rigorously prove that our formula gives the asymptotic
metric on the $N$-vortex moduli space, but the known results for 
monopoles make this conjecture
plausible. Using our formula we can calculate the scattering of two 
vortices that do not approach close to each other. The leading exponentially 
small expression for the scattering angle can be obtained exactly.

This paper is organized as follows. In Section 2 we obtain the
asymptotic 2-vortex metric, and its generalization to $N$ vortices,
along with the asymptotic Ricci and scalar curvatures. In
Section 3 we rederive the metric using the model of
vortices as point-like particles. In Section 4 we discuss the
scattering of two vortices using the asymptotic metric.

\section{Well Separated Vortices -- Field and Metric}
\setcounter{equation}{0}
\label{sec:svfm}

The key to the metric on $\M_N$, the $N$-vortex moduli space, is the
equation (\ref{heqn1}), whose
solutions determine the static $N$-vortex fields. It is sometimes convenient
to use a complex coordinate $z = x + iy$ for a general point in
the plane, and to denote the vortex locations correspondingly by
$\{ Z_r  : 1 \le r \le N \}$. Eq. (\ref{heqn1}) becomes
\be
\nabla^2 h - e^h + 1 = 4\pi \sum^N_{r=1} \delta (z - Z_r)
\label{heqn2}
\ee
where $\nabla^2 = 4 \frac{\pr^2}{\pr z \pr \bar z}$.
Around $Z_r$, the function $h(z,\bar z)$ has the local expansion
\begin{eqnarray}
h =\log|z-Z_r|^2 &+& a_r + \half\bar b_r(z - Z_r)+
\half b_r(\bar z - \bar Z_r) \nonumber \\
&+& \bar c_r(z-Z_r)^2 - \frac{1}{4}(z-Z_r)
(\bar z - \bar Z_r)+c_r(\bar z - \bar Z_r)^2 + \dots \,,
\label{hex}
\end{eqnarray}
where $a_r$ is real, and $b_r$, $c_r$ complex. Taubes proved that
this series, with the logarithmic term removed, is a convergent Taylor
expansion. The logarithmic term and the coefficient $\frac{1}{4}$ are
determined by the equation locally, but the remaining coefficients
are not. They depend on the positions of the other vortices, but not in an
explicitly known way. Most important for us is the coefficient $b_r$.

Samols' formula for the metric on $\M_N$ is
\be
\gmet =\pi \sum_{r,s=1}^N \left( \delta_{rs}+
2\frac{\pr b_s}{\pr Z_r} \right) \, dZ_r d\bar Z_s \,.
\label{met}
\ee
The functions $b_r$ obey the symmetry relation
\be
\label{symm}
\frac{\pr b_s}{\pr Z_r} = \frac{\pr \bar b_r}{\pr \bar Z_s}
\ee
and from this it follows that the metric is not only real, but also
K\"ahler. Invariance of the metric under a translation of all the vortices
implies that $\sum b_r = 0$ \cite{Sa}, and rotational invariance
implies that $\sum \bar Z_r b_r$ is real \cite{Ro}.

For well separated vortices, we assume that $h$ is exponentially small
except in a core region with radius of order 1 around each vortex, and
there $h$ has an approximate, local circular symmetry. It
follows that if the minimum separation
of any pair is $L \gg 1$, then the $\delta_{rs}$ term dominates
the metric, and the correction is of order $e^{-L}$. The
metric is therefore approximately flat.

Let us now concentrate on two vortices, and
denote their positions by
\be
Z_1 = Z + \sigma e^{i\theta} \,, \quad Z_2 = Z - \sigma e^{i\theta} \,.
\label{loc}
\ee
It follows from the symmetry of the 2-vortex field around the centre
of mass $Z$, or from the
properties of the functions $b_r$ mentioned above, that in this case $b_1 =
b(\sigma) e^{i\theta}$ and $b_2 = -b_1$, where $b(\sigma)$ is a real function.
Samols' formula implies that the moduli space metric is
\be
\gmet  = 2\pi \, dZ d\bar Z
+ \eta(\sigma) (d\sigma^2 + \sigma^2 \, d\theta^2)
\label{redmet}
\ee
where
\be
\eta(\sigma)
= 2\pi \left( 1 + \frac{1}{\sigma} \frac{d}{d\sigma} \Bigl(\sigma
b(\sigma)\Bigr) \right) \,.
\label{eta2form}
\ee
The relative motion of two vortices takes place on the reduced moduli
space where $Z$ is fixed. This is a surface of revolution.
The range of the coordinates is $0
\le \sigma < \infty$ and $-\frac{\pi}{2} \le \theta \le
\frac{\pi}{2}$, with $\theta=-\frac{\pi}{2}$ and
$\theta=\frac{\pi}{2}$ identified. The range of $\theta$
is $\pi$ and not $2\pi$ because the vortices are identical. Therefore,
the surface is asymptotically conical, rather than planar.

So far, our exposition has been a summary of known results, but now we show
how to calculate the leading asymptotic correction to the
conical metric. We return to the equation (\ref{heqn1}), and
consider first the circularly symmetric solution $h_0$ for a single
vortex at the origin. In terms of polar coordinates
$(\rho, \varphi)$, the equation satisfied by $h_0$, for $\rho > 0$, is
\be
\frac{d^2 h_0}{d\rho^2} + \frac{1}{\rho} \frac{d h_0}{d\rho} - e^{h_0}
+ 1 = 0 \,.
\label{h0eq}\label{**}
\ee
The boundary conditions are $h_0 \sim 2 \log \rho$ for small
$\rho$, and $h_0 \to 0$ as $\rho \to \infty$. The Taylor expansion of
$h_0 - 2 \log \rho$ about $\rho = 0$ involves only even powers of
$\rho$. For large $\rho$, eq. (\ref{h0eq}) has the linearized form
\be
\label{bess}
\frac{d^2 h_0}{d\rho^2} + \frac{1}{\rho} \frac{d h_0}{d\rho} - h_0 = 0
\,,
\ee
the modified Bessel equation of zeroth order, so
\beq
\label{***}
h_0(\rho) \sim \frac{q}{\pi}K_0(\rho)\, ,
\eeq
where $q$ is a constant. The corrections to
this asymptotic approximation are expected to be suppressed by order 
$e^{-\rho}$. By numerical integration of the nonlinear equation
(\ref{h0eq}), it has been determined that $q = -10.6$ \cite{Sp}.
Recently, Tong has given an argument involving dualities in string
theory which strongly suggests that $q = -2\pi \, 8^{\frac{1}{4}}$, in
agreement with the numerical result \cite{To}.
There is as yet no direct proof of this using (\ref{h0eq}).

Next, let us consider the perturbation of the solution $h_0$ due to
other, distant vortices, still assuming that one vortex, which we
label as vortex 1, is
precisely at the origin. Let us write $h = h_0 + h_1$, where $h_1$ is
small in the neighbourhood of vortex 1. The
linearization of eq. (\ref{heqn1}) implies that
\be
\left(\nabla^2 - e^{h_0}\right) h_1 = 0 \,.
\ee
The operator acting on $h_1$ has no singularity at the origin, so
$h_1$ is smooth there, and the logarithmic singularity of $h$ is
carried entirely by $h_0$. Since $h_0$ is circularly symmetric, we can
separate variables and write
\begin{eqnarray}
h(\rho, \varphi) = h_0(\rho) &+& h_1 (\rho, \varphi) \nonumber\\
:= h_0(\rho) &+& \half f_0(\rho) +
\sum_{n=1}^\infty \Bigl( f_n(\rho) \cos n\varphi
+ g_n(\rho) \sin n\varphi \Bigr) \,,
\label{h1exp}
\end{eqnarray}
where $f_n$ obeys the equation
\be
\frac{d^2 f_n}{d\rho^2} + \frac{1}{\rho} \frac{d f_n}{d\rho} -
\left(e^{h_0} + \frac{n^2}{\rho^2}\right)f_n = 0 \,,
\label{fneq}
\ee
and $g_n$ obeys the same equation. $f_n$ is nonsingular at $\rho = 0$
and has a series expansion $f_n = \alpha_n \rho^n + \dots$. Similarly,
$g_n = \beta_n \rho^n + \dots$. The expansion (\ref{h1exp}) is consistent with
the general expansion of $h$ around vortex 1, that is,
(\ref{hex}) with $r=1$ and $Z_1 = 0$.
By identifying the terms linear in $\rho$, we find that $b_1$, the
coefficient we are interested in, is given by $b_1 =
\alpha_1 + i\beta_1$.

From now on, therefore, we just consider equation (\ref{fneq}) for
$f_1$, that is,
\be
\frac{d^2 f_1}{d\rho^2} + \frac{1}{\rho} \frac{d f_1}{d\rho} -
\left(e^{h_0} + \frac{1}{\rho^2}\right)f_1 = 0 \,.
\label{f1eq}
\ee
For large $\rho$, this simplifies to
\be
\frac{d^2 f_1}{d\rho^2} + \frac{1}{\rho} \frac{d f_1}{d\rho} -
\left(1 + \frac{1}{\rho^2}\right)f_1 = 0 \,,
\label{f1asymp}
\ee
which is valid for $1 \ll \rho \ll L$, where $L$ is
the distance from the origin to the next-nearest vortex. 
Note that the difference between the coefficients in eqs. (\ref{f1eq})
and (\ref{f1asymp}) is $e^{h_0} - 1$, which is smooth, finite, and
exponentially localized. We therefore suppose that the asymptotic form of
the solutions of (\ref{f1eq}) are exact solutions of (\ref{f1asymp}). This is
supported by the results used in various
examples of scattering theory, and which follow from
Levinson's theorem \cite{Ea}; however a result of the precise type we require,
involving perturbations of Bessel's equation, appears to us not to
have been established. Eq. (\ref{f1asymp}) is the modified Bessel 
equation of first order, whose general solution is a linear combination of the
functions $K_1(\rho)$ and $I_1(\rho)$. 

Let us now assume that there is just one other vortex, vortex 2,
whose location (in Cartesian coordinates) is $(-2\sigma,0)$, with
$\sigma \gg 1$. In this case,
$h$ is reflection symmetric under $\varphi \mapsto -\varphi$, so in
the Fourier series for $h_1$ all the functions $g_n$
vanish. In particular, $b_1 = \alpha_1$, and is real.

In the region far from both vortex centres, the equation (\ref{heqn1})
linearizes to
\be
\nabla^2 h - h = 0
\label{helm}
\ee
and is solved by the linear superposition of the fields due
to each vortex separately
\be
h(\rho, \varphi) = \frac{q}{\pi} K_0(\rho) + \frac{q}{\pi}
K_0 \left(\sqrt{4\sigma^2 + 4\sigma\rho\cos\varphi + \rho^2}\right) \,.
\label{super}
\ee
The argument of the second $K_0$ function is the distance to vortex 2
from the point with polar coordinates $(\rho, \varphi)$.
By separation of variables, the general solution of the Helmholtz
equation (\ref{helm}), regular at $\rho = 0$ and with the reflection
symmetry $\varphi \mapsto -\varphi$, is a linear combination of the functions
$I_n(\rho) \cos n\varphi$. The function
$K_0 \left(\sqrt{4\sigma^2 + 4\sigma\rho\cos\varphi +
\rho^2}\right)$ is such a solution (whereas $K_0(\rho)$ is not, being
singular at $\rho = 0$), so
\be
K_0\left(\sqrt{4\sigma^2 + 4\sigma\rho\cos\varphi + \rho^2}\right)
= k_0 I_0(\rho) + 2\sum_{n=1}^\infty k_n I_n (\rho) \cos(n\varphi)
\label{K0expan}
\ee
for some real constants $k_n$.

Note that an important special solution of (\ref{helm}) is $e^x =
e^{\rho\cos\varphi}$, and its expansion
\be
e^{\rho\cos\varphi} = I_0(\rho)
+ 2\sum_{n=1}^\infty I_n(\rho) \cos(n\varphi)
\ee
defines the functions $I_n(\rho)$. Combining the series for the
exponential function with trignometric identities, one can compute the
leading terms in the series expansions for $I_n$. These are also given in
standard references, e.g. \cite{Er}. It is sufficient for us to
record that
\be
I_0(\rho) = 1 + \dots \,, \quad I_1(\rho) = \half \rho + \dots \,.
\label{I0I1}
\ee

We can now return to (\ref{K0expan}) and determine
$k_1$, the coefficient we need. The Taylor expansion (in $\rho$)
of the two sides gives
\be
K_0(2\sigma) - K_1(2\sigma)\rho \cos \varphi + \dots
= k_0 + k_1 \rho \cos \varphi + \dots \,,
\label{Tay}
\ee
where we have used the identity $K_1 = -K_0'$, and
the results above for $I_0$ and $I_1$. So $k_0 = K_0(2\sigma)$ and
\be
k_1 = -K_1(2\sigma) \,.
\ee

With this result, we can now match the Fourier expansion of (\ref{super}),
the linearized field $h$ due to the
two vortices, valid outside their cores, with the Fourier expansion of
$h = h_0 + h_1$ near vortex 1. In the range $1 \ll \rho \ll 2\sigma$, we
find
\begin{eqnarray}
\frac{q}{\pi}\Bigl( K_0(\rho) + K_0(2\sigma)I_0(\rho) \Bigr)
&-& \frac{2q}{\pi} K_1(2\sigma) I_1(\rho)\cos\varphi + \dots \nonumber\\
&=& \frac{q}{\pi} K_0(\rho) + \half f_0(\rho)
+ f_1(\rho)\cos\varphi + \dots \label{match}
\end{eqnarray}
Therefore, $f_1(\rho)$ has the asymptotic form
\be
f_1(\rho) = -\frac{2q}{\pi} K_1(2\sigma) I_1(\rho)
\ee
and there is no $K_1(\rho)$ piece. The further terms on both sides
of (\ref{match}) involve $\cos n\varphi$ with $n>1$, and
could be determined from higher order terms in the Taylor
expansion (\ref{Tay}), but we do not need these.

The last step is to extrapolate the function $f_1$
into the core region of vortex 1. It is rather remarkable that this can be
done, because the equation satisfied by $f_1$, namely (\ref{f1eq}),
is not a standard equation, and the coefficient $e^{h_0}$ is not known
explicitly. However, one solution of (\ref{f1eq}) is known. It is
\be
\tilde f_1 = \frac{dh_0}{d\rho} \,.
\ee
This can be verified by differentiating (\ref{h0eq}), the nonlinear
equation for $h_0$. The interpretation of this solution is that it
corresponds to the translational zero mode of vortex 1. If the centre
of that vortex is infinitesimally translated by $\epsilon$ in the
$x$-direction, then the field of vortex 1 becomes
$h = h_0(\rho - \epsilon \cos \varphi)$ to first order
in $\epsilon$. So $h= h_0 + h_1$ where
$h_1 = -\epsilon\tilde f_1 \cos \varphi$ and
$\tilde f_1 = \frac{dh_0}{d\rho}$. Since $h_0 =
\frac{q}{\pi}K_0(\rho)$ asymptotically, it follows that $\tilde f_1 =
-\frac{q}{\pi}K_1(\rho)$ asymptotically. Similarly, since $h_0 \sim
2\log \rho$ for small $\rho$, it follows that $\tilde f_1 \sim 2/\rho$
for small $\rho$. By contrast, the solution $f_1$ of (\ref{f1eq}) that really
interests us has the asymptotic behaviour $-\frac{2q}{\pi} K_1(2\sigma)
I_1(\rho)$ for large $\rho$, and the finite linear behaviour $b_1 \rho$
for small $\rho$, where $b_1$ is to be found.

Now, equation (\ref{f1eq}) has a Wronskian identity
\be
\rho \left( \tilde f_1 \frac{df_1}{d\rho} - f_1 \frac{d \tilde
f_1}{d\rho} \right) = {\rm constant} \,,
\label{Wron}
\ee
relating the two solutions $f_1$ and $\tilde f_1$. Using the
asymptotic forms of $f_1$ and $\tilde f_1$, and the
Wronskian identity for the modified Bessel functions
\be
\rho \left( K_1 \frac{dI_1}{d\rho} - I_1 \frac{d K_1}{d\rho} \right) =
1 \,,
\ee
we deduce that the constant in (\ref{Wron}) is
$\frac{2q^2}{\pi^2}K_1(2\sigma)$. Evaluating (\ref{Wron}) near $\rho
= 0$ we deduce, finally, that
\be
b_1 = \frac{q^2}{2\pi^2}K_1(2\sigma) \,.
\label{basymp}
\ee

We can now use this result to calculate the 2-vortex metric. In the
above calculation, vortex 1 was at
the origin and vortex 2 at $(-2\sigma,0)$. From (\ref{loc}) we see
that $Z=-\sigma$ and $\theta = 0$, so
\be
b(\sigma) = \frac{q^2}{2\pi^2}K_1(2\sigma) \,.
\ee
Therefore the prefactor $\eta$ in the
metric (\ref{redmet}) is
\be
\eta(\sigma) = 2\pi \left( 1 - \frac{q^2}{\pi^2}K_0(2\sigma) \right) \,,
\label{eta}
\ee
where we have used (\ref{eta2form}) and the identity
$K_1'(s) + K_1(s)/s = - K_0(s)$. The complete asymptotic 2-vortex
metric is
\be
\label{asymet}
\gmet  = 2\pi \, dZ d\bar Z
+ 2\pi \left( 1 - \frac{q^2}{\pi^2}K_0(2\sigma) \right)
(d\sigma^2 + \sigma^2 \, d\theta^2) \,.
\label{asymp2}
\ee
We shall investigate the geodesics of this metric in section 4, and
hence determine how vortices scatter.

To extend (\ref{asymp2}) to the asymptotic $N$-vortex metric is not hard.
Let us use the complex coordinates of the vortices $Z_r$ and introduce
the notation $Z_{rs} := Z_r - Z_s$.
The flat part of the metric (\ref{met}) can be reexpressed as
\be
\pi \sum_{r=1}^N dZ_r d\bar Z_r = N\pi dZ d\bar Z + \frac{\pi}{2N}
\sum _{r \ne s} dZ_{rs} d\bar Z_{rs} \,,
\ee
where $Z$ is the centre of mass coordinate
\be
\label{Zdef}
Z = \frac{1}{N} (Z_1 + Z_2 + \dots + Z_N) \,.
\ee
Note that the differentials $dZ_{rs}$ are not all linearly
independent.

To find the remaining part of the metric, we need to find $b_s$ and
its derivatives. The solution of the Helmholtz equation (\ref{helm}) becomes a
linear superposition of the fields due to the $N$ vortices. The asymptotic
matching of $h$ in the neighbourhood of the $s$-th vortex can be carried
out as before. This leads to the following expression for $b_s$ that
is a linear superposition of the effects of the other $N-1$ vortices,
\be
b_s = \frac{q^2}{2\pi^2}\sum_{r \ne s} K_1(|Z_{sr}|)
\frac{Z_{sr}}{|Z_{sr}|} \,.
\label{bsasymp}
\ee
Each term is the obvious generalization of (\ref{basymp}), combined with
the orientational phase factor $Z_{sr}/|Z_{sr}|$ which reduces to
$e^{i\theta}$ for two vortices.

Because of translational invariance,
\be
\sum_{r=1}^N \frac{\pr b_s}{\pr Z_r} = 0 \,,
\ee
so
\be
\frac{\pr b_s}{\pr Z_s} = -\sum_{r \ne s} \frac{\pr b_s}{\pr Z_r}
\label{dbsdZs}
\ee
(no summation over $s$). For $r \ne s$, we find,
differentiating (\ref{bsasymp}) with respect to $Z_r$ and keeping
$\bar Z_r$ fixed, that
\be
\frac{\pr b_s}{\pr Z_r} = \frac{q^2}{4\pi^2} K_0(|Z_{sr}|) \,.
\label{dbsdZr}
\ee
Eq. (\ref{dbsdZr}) combined with (\ref{dbsdZs}) gives
\be
\sum_{r,s=1}^N \frac{\pr b_s}{\pr Z_r} \, dZ_r d\bar Z_s
=\frac{q^2}{4\pi^2} \sum_{r \ne s} K_0(|Z_{sr}|) \, (dZ_r - dZ_s)
\, d\bar Z_s \,.
\ee
Since $K_0(|Z_{sr}|)=K_0(|Z_{rs}|)$, we symmetrize over the contributions
of these two terms, obtaining
\be
\sum_{r,s=1}^N \frac{\pr b_s}{\pr Z_r} \, dZ_r d\bar Z_s
=-\frac{q^2}{8\pi^2} \sum_{r \ne s} K_0(|Z_{rs}|) \, dZ_{rs} d\bar
Z_{rs} \,.
\ee

Putting these ingredients together, we obtain our final expression for the
asymptotic $N$-vortex metric
\be
\label{gN}
\gmet  = N\pi \, dZ d\bar Z + \pi \sum_{r \ne s} \left( \frac{1}{2N} -
\frac{q^2}{4\pi^2} K_0(|Z_{rs}|) \right) \, dZ_{rs} d\bar Z_{rs} \,.
\label{asympN}
\ee
For two vortices, located at the points (\ref{loc}), this reduces to
(\ref{asymp2}). Since the coefficients in the asymptotic metric
depend only on the magnitudes of the vortex separations, it is clear
that it is translationally and rotationally symmetric. The structure of the
metric as a small perturbation of the flat Euclidean metric becomes
clearer if we eliminate the centre of mass coordinate $Z$ using (\ref{Zdef}):
\beq
\label{gN2}
\gmet =\pi\sum_r dZ_r d\bar Z_r-\frac{q^2}{4\pi}\sum_{r\neq
s}K_0(|Z_{rs}|) \, dZ_{rs} d\bar Z_{rs} \,.
\eeq

One way to see that this metric is K\"ahler is to note that eq. (\ref{symm}) is
satisfied, since (\ref{dbsdZr}) and (\ref{dbsdZs}) imply that
$\frac{\pr b_s}{\pr Z_r}$ is real and symmetric. More explicitly, the
asymptotic K\"ahler form is
\beq
\omega=\frac{iN\pi}{2} dZ\wedge d\bar {Z}
+ \frac{i\pi}{2}\sum_{r\neq s}\left( \frac{1}{2N} -
\frac{q^2}{4\pi^2} K_0(|Z_{rs}|) \right) dZ_{rs}\wedge d\bar Z_{rs} \,.
\label{gN3}
\eeq
Since the 1-forms $dZ$, $dZ_{rs}$ are closed, one finds that
\beq
d\omega=-\frac{iq^2}{16\pi}\sum_{r\neq s}
\frac{K_0'(|Z_{rs}|)}{|Z_{rs}|}\left(
\bar{Z}_{rs} \, dZ_{rs}\wedge dZ_{rs}\wedge d\bar Z_{rs}+
Z_{rs} \, d\bar Z_{rs}\wedge dZ_{rs}\wedge d\bar Z_{rs}\right)=0 \,,
\eeq
so $\omega$ is closed. The K\"ahler
potential is
\be
\pi \sum_{r=1}^N Z_r \bar Z_r - \frac{q^2}{\pi}
\sum_{r \ne s} K_0(|Z_{rs}|) \,.
\ee
The K\"ahler form is of direct interest in certain
non-relativistic models of vortex dynamics \cite{Ro}. Such models have first
order dynamics in time, and it is conjectured that slow vortex dynamics is
well approximated by a Hamiltonian flow on the $N$-vortex moduli
space, where the symplectic structure is precisely this K\"ahler form.
Clearly the closure of $\omega$ is crucial for this to make sense.

The curvature properties of soliton moduli spaces are of some interest. 
For example, the scalar curvature of $\M_N$ is relevant to 
quantum $N$-soliton dynamics \cite{mosshi}, 
while in the case of monopoles, 
Ricci flatness of $\M_N$ was the key property exploited in Atiyah and 
Hitchin's construction of the metric for $N=2$. In order to compute the
asymptotic Ricci tensor for the $N$-vortex metric (\ref{gN2}), it is 
convenient to write $g$ as
\beq\label{r1}
g=\sum_{r,s} g_{rs}dZ_r d\bar Z_s=
\sum_{r,s}\pi(\delta_{rs}+h_{rs})dZ_r d\bar Z_s \,,
\eeq
and work up to linear order in the perturbation $h$. It is a standard result
in K\"ahler geometry \cite{wil}
that the Ricci tensor associated with $g$ is
\beq
\label{kaeric}
\ric=-\sum_{r,s}\frac{\cd^2\log G}{\cd Z_r\cd\bar Z_s}\, d Z_r d\bar Z_s
\eeq
where $G$ is the determinant of the hermitian coefficient matrix $g_{rs}$.
In this case,
\bea
G&=&\det\pi(\I+h)=\pi^N(1+\tr\, h+\cdots)\nonumber\\
\Rightarrow\quad
\log G&=&N\log\pi+\sum_r h_{rr}+\cdots\nonumber\\
&=&N\log\pi-\frac{q^2}{2\pi^2}\sum_{r\neq s}K_0(|Z_r-Z_s|)\label{r2}
\eea
from (\ref{gN2}). Equations (\ref{kaeric}) 
and (\ref{r2}) together with Bessel's
equation imply that
\beq
\ric=\frac{q^2}{8\pi^2}\sum_{r\neq s}K_0(|Z_r-Z_s|)dZ_{rs}d\bar Z_{rs}\, .
\eeq
One sees that the $N$-vortex Ricci tensor is asymptotically positive
semi-definite, its two-dimensional null space being tangent to the
translation orbits in $\M_N$ (that is, a vector is null if and only if
it generates a rigid translation of the $N$-vortex system). 
Tracing $\ric$,
one obtains
the scalar curvature,
\beq
\scal=\sum_{r,s}g^{rs}\ric_{rs}=\frac{q^2}{4\pi^3}\sum_{r\neq s}
K_0(|Z_r-Z_s|)+\cdots\, ,
\eeq
whence one sees that $\M_N$ is asymptotically scalar positive. It is an
interesting open question whether the true metric on $\M_N$ has similar
curvature positivity properties. The numerical results of Samols
for $N=2$ suggest that it may \cite{Sa}.

\section{The Point Source Formalism}
\label{sec:psf}
\setcounter{equation}{0}

In this section we rederive the asymptotic 2-vortex metric from a more
physical viewpoint. The idea is that, viewed from afar, a static vortex
looks like a solution of a linear field theory with a
point source at the vortex centre. We will see that the appropriate point
source is a composite scalar monopole and magnetic dipole in a
Klein-Gordon/Proca theory. If physics is to be model independent, the
forces between vortices should approach those between the corresponding
point particles in the linear theory as their separation grows. This idea,
which originated in the context of monopole dynamics \cite{Ma2}, has
already been successfully used to obtain an asymptotic formula for
{\em static} intervortex forces away from critical coupling \cite{Sp}. The
present application is somewhat more subtle since we are required to
analyze the interaction between point sources moving along arbitrary
trajectories. We handle the problem perturbatively: using a mixture of
Lorentz invariance and conservation properties we obtain expressions for a
moving point source and the Klein-Gordon/Proca field it induces correct up
to acceleration terms. From these we construct the interaction Lagrangian
for one moving point source interacting with the field induced by another.
This Lagrangian is purely kinetic, i.e. quadratic in velocities, and hence may
naturally be reinterpreted as the energy associated with geodesic flow on
the asymptotic 2-vortex moduli space. The extension to $N$-vortex
dynamics is entirely trivial.

In this section $x^\mu = (x^0,x^1,x^2)$ denotes a space-time
point. $x^0=t$ is the time and $\xv = (x^1,x^2)$ denotes a spatial point.
To linearize the abelian Higgs theory (\ref{abHig}), we choose the
gauge so that the scalar
field $\phi$ is real. Since vortices have nontrivial winding at infinity,
this requires a gauge transformation which is singular at the vortex centre.
This need not concern us since we seek only to replicate the local, far field
behaviour of the vortex in the linear theory. In this gauge, the vacuum is
$\phi=1$, so we define $\phi=1+\psi$ and linearize in $\psi$. The
resulting Lagrangian density is
\beq
{\cal L}=\half\cd_\mu\psi\cd^\mu\psi-\half\psi^2-\frac{1}{4}F_{\mu\nu}
F^{\mu\nu}+\half A_\mu A^\mu+\kappa\psi-j^\mu A_\mu
\eeq
where $\kappa$ is the scalar charge density and $j$ the electromagnetic
current density. These will be chosen to replicate the vortex asymptotics.
The corresponding field equations are
\bea
\label{kg}
(\Box+1)\psi&=&\kappa \\
\label{proca}
(\Box+1)A^\mu&=&j^\mu
\eea
where $\Box=\cd_\mu\cd^\mu=\cd_t^2-\nabla^2$, and we assume that $j$ is a
conserved current, $\cd_\mu j^\mu=0$.

In this gauge, the scalar field of a single vortex located at the origin is
$\phi=\exp(\half h_0)$ where $h_0$ satisfies (\ref{**}), so
\beq
\phi=1+\half h_0+\ldots\sim 1+\frac{q}{2\pi}K_0(|\xv|)
\eeq
for large $|\xv|$ by eq.\ (\ref{***}). Hence we seek a point source $\kappa$
so that the solution of (\ref{kg}) is $\psi=\frac{q}{2\pi} K_0(|\xv|)$.
Since the static Klein-Gordon equation (Helmholtz equation) in two
dimensions has Green's function $K_0$,
\beq
\label{gf}
(-\nabla^2+1)K_0(|\xv|)=2\pi\delta(\xv) \,,
\eeq
one sees that
\beq
\label{scalso}
\kappa=q\delta(\xv) \,.
\eeq
For a static vortex, the time component of the gauge potential $A_0$
vanishes. The asymptotic behaviour
of its spatial components $A_i$ is determined by the first Bogomolny
equation (\ref{Bog1}) which, on linearization, implies
\beq
\cd_1\psi-A_2+i(\cd_2\psi+A_1)=0 \,.
\eeq
Hence
\beq
\Av=(A^1,A^2)=(\cd_2,-\cd_1)\psi=-\frac{q}{2\pi}\kv\times\nabla K_0(|\xv|)
\eeq
where we have introduced $\kv$, the unit vector in a fictitious
$x^3$-direction orthogonal to the physical plane. It follows that the point
source which reproduces the asymptotic vortex gauge field in eq.\
(\ref{proca}) is
\beq
\label{vecso}
(j^0,\jv)=(0,-q\kv\times\nabla\delta(\xv)) \,.
\eeq
The physical interpretation of (\ref{scalso}) and (\ref{vecso}) is
that the point
particle corresponding to a single vortex at rest is a composite consisting
of a scalar monopole of charge $q$ and a magnetic dipole of moment $q$. We
shall refer to this composite as a (static) point vortex.

The interaction between two arbitrary (possibly time-dependent) composite
sources $(\kappa_{(1)},j_{(1)})$ and $(\kappa_{(2)},j_{(2)})$ in this
linear theory is described by the Lagrangian
\beq
\label{lint}
\Li=\int d^2\xv\, (\kappa_{(1)}\psi_{(2)}-j^\mu_{(1)}A_{(2)\mu})
\eeq
where $(\psi_{(i)},A_{(i)})$ are the fields induced by
$(\kappa_{(i)},j_{(i)})$ according to the wave equations
(\ref{kg}), (\ref{proca}).
This is obtained by extracting the cross terms in $\int d^2\xv\,
 {\cal L}$
where $(\kappa,j)=(\kappa_{(1)},j_{(1)})+(\kappa_{(2)},j_{(2)})$ and
$(\psi,A)=(\psi_{(1)},A_{(1)})+(\psi_{(2)},A_{(2)})$ by linearity.
Although (\ref{lint}) looks asymmetric under interchange of sources, it is
not, as may be shown using (\ref{kg}), (\ref{proca}) and integration
by parts. If
the sources are chosen to be static point vortices, that is, translated
versions of (\ref{scalso}) and (\ref{vecso}), one finds that $\Li=0$,
so static point
vortices exert no net force on one another at critical coupling, in
agreement with the nonlinear theory.

We seek to compute $\Li$ in the case where the two sources represent
point vortices moving along arbitrary trajectories in $\R^2$. To do so,
we must construct a time-dependent point source representing a vortex
moving along some curve $\yv(t)$, say. The construction is guided by two
principles: first, in the case of motion at constant velocity, the source
should reduce to (\ref{scalso}), (\ref{vecso}) in the vortex's rest frame;
second, for any trajectory the vector source $j$, which represents the
vortex's electromagnetic current density, must remain a conserved current.
The result will be correct up to quadratic order in velocity and linear
order in acceleration.

It is straightforward to calculate Lorentz boosted versions of the sources
(\ref{scalso}), (\ref{vecso}). Let $\xi^\mu$ denote the rest frame coordinates
and assume that at time $t=0$, the point vortex lies at the origin,
$\xv=0$, and is moving with velocity $\uv$. Then, by decomposing $\xv$ and
$\xivec$ into their components parallel and perpendicular to $\uv$, one
finds that at this time, $\xi^\prl=\gamma(u)x^\prl$ (where $\gamma(u)=
(1-u^2)^{-\half}$ is the usual contraction factor)
while $\xi^\perp=x^\perp$. Hence
\beq
\label{xx0}
\xivec=\gamma(u)\frac{\xv\cdot\uv}{u}\, \uv+\left(\xv-
\frac{\xv\cdot\uv}{u}\, \uv\right)=\xv+\half(\xv\cdot\uv)\uv+\ldots
\eeq
where the ellipsis denotes discarded terms of order $u^4$ or greater. We
shall not persist in so denoting these terms. Rather, we shall include an
ellipsis only where further neglible terms have been dropped to obtain the
given expression. Since $\kappa$ is a Lorentz scalar, the boosted scalar
monopole is $\kappa(\xv)=\kappa_{{\rm static}}(\xivec)=
q\delta(\xivec)$. To interpret this delta function as a
distribution on the $\xv$-plane, note that for any test function $f$,
\beq
\int d^2\xv\, f(\xv)\delta(\xivec)=\int d^2\xivec\, \left|\frac{\cd\xv}{\cd
\xivec}\right|f(\xv(\xivec))\delta(\xivec)\nonumber \\
=(1-\half u^2)f(\zerovec)+\ldots \,.
\eeq
Therefore $\delta(\xivec)=(1-\half u^2)\delta(\xv)$ and so, at $t=0,$
\beq
\label{s1}
\kappa(\xv)=q(1-\half u^2)\delta(\xv) \,.
\eeq

The boosted dipole is more subtle, since $j$ itself transforms as a Lorentz
vector. The rest frame source is
\beq
j_{{\rm static}}=(j^0_{(0)},\jv_{(0)})=
(0,-q\kv\times\nabla_\xi\delta(\xivec)) \,.
\eeq
To obtain the laboratory frame source, one must perform a Lorentz boost on
this with velocity $-\uv$. Explicitly,
\bea
j^0(\xv)&=&u\gamma(u)\, j_{(0)}^\prl(\xivec(\xv))=\uv\cdot
\jv_{(0)}(\xivec(\xv))+\ldots\label{xx1} \\
j^\prl(\xv)&=&\gamma(u)j_{(0)}^\prl(\xivec(\xv))=
(1+\half u^2)j_{(0)}^\prl(\xivec(\xv))+\ldots\label{xx2} \\
j^\perp(\xv)&=&j^\perp_{(0)}(\xivec(\xv)) \,.
\label{xx3}
\eea
We may combine (\ref{xx2}) and (\ref{xx3}) into a single equation for
$\jv$ by using the same polarization trick  as in (\ref{xx0}), yielding
\beq
\label{xx4}
\jv(\xv)=\jv_{(0)}(\xivec(\xv))+\half(\jv_{(0)}(\xivec(\xv))\cdot\uv)\,
\uv \,.
\eeq
Now
\beq
\label{xxx}
\nabla_\xi\delta(\xivec)=\left(\frac{\cd\xv}{\cd\xivec}\right)\nabla
(1-\half u^2)\delta(\xv)=\left[(1-\half u^2)\nabla-\half\uv\,
(\uv\cdot\nabla)\right]\delta(\xv)+\ldots \,.
\eeq
Hence
\bea
j^0(\xv)&=&q(\kv\times\uv)\cdot\nabla\delta(\xv)+\ldots\label{s2}\\
\jv(\xv)&=&-q(1-\half u^2)\kv\times\nabla\delta(\xv)+\half q[\uv\,
(\kv\times\uv)\cdot\nabla+(\kv\times\uv)\, \uv\cdot\nabla]\delta(\xv)\nonumber
\\
&=&-q\kv\times\nabla\delta(\xv)+q(\kv\times\uv)\, \uv\cdot\nabla\delta(\xv) \,.
\label{s3}
\eea

By replacing $\xv$ by $\xv-\yv(t)$ and $\uv$ by $\yd(t)$ in
(\ref{s1}), (\ref{s2}) and (\ref{s3}) we obtain expressions for
the instantaneously
Lorentz boosted point vortex travelling along an arbitrary trajectory. In
particular,
\beq
\label{sb2}
j_{\rm boost}=q((\kv\times\yd)\cdot\nabla,-\kv\times\nabla+(\kv\times\yd)\,
\yd\cdot\nabla)\delta(\xv-\yv) \,.
\eeq
Note that
\beq
\cd_\mu j^\mu_{\rm boost}=q(\kv\times\ydd)\cdot\nabla\delta(\xv-\yv)+\ldots
\neq 0
\eeq
so $j_{\rm boost}$ is {\em not} the moving point source we seek. Rather, we
must add a correction $j_{\rm acc}$ to $j_{\rm boost}$ of order $\ydd$ to
enforce current conservation. Such a term will vanish in the case of
motion at constant
velocity and hence does not conflict with the required Lorentz
properties of $j$. We make the simplest choice, namely
\beq
j_{\rm acc}=(0,-q\kv\times\ydd\, \delta(\xv-\yv)) \,.
\eeq
Though we have chosen $j^0_{\rm acc}=0$,
any function of order $|\ydd|$ would do since $\cd_0 j^0_{\rm acc}$
is automatically of negligible order.
It turns out that this ambiguity has no bearing on our calculation because
$j^0_{\rm acc}$ makes no contribution to $\Li$ at order $(\mbox{velocity})^2$.

To summarize, the point vortex moving along a trajectory $\yv(t)$ is
represented by a composite point source
\bea
\label{st1}
\kappa(t,\xv)&=&q(1-\half|\yd|^2)\delta(\xv-\yv)\\
\label{st2}
j(t,\xv)&=&q\left((\kv\times\yd)\cdot\nabla,-\kv\times\nabla+
(\kv\times\yd)\, \yd\cdot\nabla-\kv\times\ydd\right)\delta(\xv-\yv) \,.
\eea

The second task in the calculation of $\Li$ is to construct the fields
$(\psi_{(2)},A_{(2)})$ induced by the second moving vortex. Were the field
equations (\ref{kg}) and (\ref{proca}) massless, we could simply use retarded
potentials, making a suitable expansion in time derivatives. Since they
are not massless, we need a substitute for this procedure. We handle the
problem by introducing formal temporal Fourier transforms of the fields
and sources, as follows. Let $\psi(t,\xv)$ be the field induced by the
time varying source $\kappa(t,\xv)$ according to (\ref{kg}), and define
Fourier transforms $\wt{\psi}$ and $\wt{\kappa}$ with variable $\omega$
dual to $t$,
\beq
\label{f0}
\psi(t,\xv):=\int_{-\infty}^{\infty}d\omega\, e^{i\omega t}\wt{\psi}(\omega,t) \,,
\qquad
\kappa(t,\xv):=\int_{-\infty}^{\infty}d\omega\,
e^{i\omega t}\wt{\kappa}(\omega,t) \,.
\eeq
Then (\ref{kg}) implies
\beq
[-\nabla^2+(1-\omega^2)]\wt{\psi}=\wt{\kappa} \,,
\label{f1}
\eeq
so for each value of $\omega$, $\wt{\psi}(\omega,\cdot)$
satisfies the static inhomogeneous Klein-Gordon equation with
mass $\sqrt{1-\omega^2}$ and source $\wt{\kappa}(\omega,\cdot)$. Comparing
with (\ref{gf}) one sees that (\ref{f1}) is solved, at least formally,
by convolution of $\wt{\kappa}(\omega,\cdot)$ with the (suitably scaled)
Green's function $K_0$,
\beq
\wt{\psi}(\omega,\xv)=\frac{1}{2\pi}\int d^2\xv'
K_0\left(\sqrt{1-\omega^2}|\xv-\xv'|\right)\wt{\kappa}(\omega,\xv') \,.
\eeq
Expanding the Green's function in $\omega$ and truncating at order $\omega^2$
yields
\beq
\wt{\psi}(\omega,\xv)=\frac{1}{2\pi}\int d^2\xv'\left[
K_0(|\xv-\xv'|)-\frac{\omega^2}{2}|\xv-\xv'|K_0'(|\xv-\xv'|)\right]
\wt{\kappa}(\omega,\xv') \,.
\eeq
From this we obtain $\psi$ by (\ref{f0}),
\beq
\label{retpot1}
\psi(t,\xv)=\frac{1}{2\pi}\int d^2\xv'\left[K_0(|\xv-\xv'|)\kappa(t,\xv')
+\Upsilon(|\xv-\xv'|)\cd_t^2\kappa(t,\xv')\right]
\eeq
where we have defined
\beq
\label{ups}
\Upsilon(s):=\half sK_0'(s) \,.
\eeq
Note that truncating the expansion in $\omega$ is, in effect, the same as
neglecting higher time derivatives of $\kappa$, and hence eventually
of $\yv(t)$ in our application. No claim of rigour is attached to the
above Fourier transform manoeuvre. One should regard it as a convenient
algebraic shorthand for generating a perturbative solution of (\ref{kg}).
Direct substitution of (\ref{retpot1}) into (\ref{kg}) confirms that this
really is a solution up to higher time derivative terms ($\cd_t^3\kappa$
etc.). Since equation (\ref{proca}) is formally identical to (\ref{kg}), the
vector field induced by a time dependent source $j$ is
\beq
\label{retpot2}
A^\mu(t,\xv)=\frac{1}{2\pi}\int d^2\xv'\left[K_0(|\xv-\xv'|)j^\mu(t,\xv')+
\Upsilon(|\xv-\xv'|)\cd_t^2j^\mu(t,\xv')\right]+\ldots \,.
\eeq

Having obtained $(\psi_{(2)},A_{(2)})$ induced by the time varying source
$(\kappa_{(2)},j_{(2)})$ we may compute the Lagrangian governing its
interaction with another source $(\kappa_{(1)},j_{(1)})$ by substitution
of (\ref{retpot1}) and (\ref{retpot2}) into (\ref{lint}). The result is
\bea
\Li&=&\frac{1}{2\pi}\int d^2\xv\, d^2\xv'\left\{K_0(|\xv-\xv'|)\left[
\kappa_{(1)}(t,\xv)\kappa_{(2)}(t,\xv')-j^\mu_{(1)}(t,\xv)j_{(2)\mu}(t,\xv')
\right]\right.\nonumber \\
\label{lint2}
&&\left.\qquad\qquad-
\Upsilon(|\xv-\xv'|)\left[\cd_t\kappa_{(1)}(t,\xv)
\cd_t\kappa_{(2)}(t,\xv')-\cd_t j^\mu_{(1)}(t,\xv)\cd_t
j_{(2)\mu}(t,\xv')\right]\right\}
\eea
where a total time derivative has been discarded. It remains to substitute
the point vortex sources (\ref{st1}), (\ref{st2}) for vortices moving along
trajectories $\yv(t)$ and $\zv(t)$ into (\ref{lint2}) and evaluate the
integrals. Explicitly,
\bea
\Li&=&\frac{1}{2\pi}\int d^2\xv\, d^2\xv'[\kappa_{(1)}\kappa_{(2)}+
\jv_{(1)}\cdot\jv_{(2)}-j^0_{(1)}j^0_{(2)}]K_0(|\xv-\xv'|)\nonumber \\
&&-\frac{1}{2\pi}\int d^2\xv\, d^2\xv'[\cd_t\kappa_{(1)}\cd_t\kappa_{(2)}+
\cd_t\jv_{(1)}\cdot\cd_t\jv_{(2)}-\cd_tj^0_{(1)}\cd_tj^0_{(2)}]
\Upsilon(|\xv-\xv'|)
\eea
where
\bea
\kappa_{(1)}\kappa_{(2)}&=& q^2[1-\half(|\yd|^2+|\zd|^2)]\delta(\xv-\yv)
\delta(\xv'-\zv) \nonumber \\
\jv_{(1)}\cdot\jv_{(2)}&=& q^2[\nabla_y\cdot\nabla_z-
(\zd\cdot\nabla_y)(\zd\cdot\nabla_z)-(\yd\cdot\nabla_y)(\yd\cdot\nabla_z)
\nonumber \\
&&\qquad\qquad\qquad
-\zdd\cdot\nabla_y-\ydd\cdot\nabla_z]\delta(\xv-\yv)\delta(\xv'-\zv)
\nonumber \\
j^0_{(1)}j^0_{(2)}&=& q^2((\kv\times\yd)\cdot\nabla_y) 
((\kv\times\zd)\cdot\nabla_z) \delta(\xv-\yv)\delta(\xv'-\zv) \nonumber \\
\cd_t\kappa_{(1)}\cd_t\kappa_{(2)}&=& q^2(\yd\cdot\nabla_y)
(\zd\cdot\nabla_z)\delta(\xv-\yv)\delta(\xv'-\zv) \nonumber \\
\cd_t\jv_{(1)}\cdot\cd_t\jv_{(2)}&=& q^2(\yd\cdot\nabla_y)
(\zd\cdot\nabla_z)\nabla_y\cdot\nabla_z\, 
\delta(\xv-\yv)\delta(\xv'-\zv)\nonumber \\
\cd_tj^0_{(1)}\cd_tj^0_{(2)}&=&0 \,.
\eea
We have discarded terms of order $|\yd|^3$, $|\ydd||\zd|$ etc., and
systematically used
\beq
\nabla\delta(\xv-\yv)=-\nabla_y\delta(\xv-\yv),\qquad
\nabla'\delta(\xv'-\yv)=-\nabla_z\delta(\xv'-\zv) \,.
\eeq
All the integrals are thus rendered trivial, so
\bea
\Li&=&\frac{q^2}{2\pi}\Bigl\{[1-\half(|\yd|^2+|\zd|^2)+\nabla_y\cdot
\nabla_z-(\zd\cdot\nabla_y)(\zd\cdot\nabla_z)-
(\yd\cdot\nabla_y)(\yd\cdot\nabla_z)
\nonumber \\
&&\qquad-\zdd\cdot\nabla_y-\ydd\cdot\nabla_z-((\kv\times\yd)\cdot\nabla_y)
((\kv\times\zd)\cdot\nabla_z)]K_0(|\yv-\zv|)\nonumber \\
&&\qquad-(\yd\cdot\nabla_y)(\zd\cdot\nabla_z)[1+\nabla_y\cdot\nabla_z]
\Upsilon(|\yv-\zv|)\Bigr\} \,.
\label{lint3}
\eea
We now use the following identities, which are easily checked using
Bessel's equation (\ref{bess}):
\bea
\label{ia}
(1+\nabla_y\cdot\nabla_z)K_0(|\yv-\zv|)&=&0\\
\label{ib}
(1+\nabla_y\cdot\nabla_z)\Upsilon(|\yv-\zv|)&=&-K_0(|\yv-\zv|)\\
(\alvec\cdot\nabla_y)(\betvec\cdot\nabla_z)K_0(|\yv-\zv|)&=&
-\frac{\alvec\cdot\betvec\, K_0'(|\yv-\zv|)}{|\yv-\zv|}-
\alpha^\prl\beta^\prl\Lambda(|\yv-\zv|) \,.
\label{ic}
\eea
In (\ref{ic}), $\alvec$, $\betvec$ are any vectors with $\betvec$ independent
of $\yv$, and $\Lambda$ denotes the function
\beq
\label{lam}
\Lambda(s):=K_0(s)-2\frac{K_0'(s)}{s} \,.
\eeq
Also, we have used the decomposition of $\alvec$, $\betvec$ relative to
the orthonormal frame
\beq
\nv^\prl=\frac{\yv-\zv}{|\yv-\zv|},\qquad
\nv^\perp=\kv\times\nv^\prl \,.
\eeq
Equations (\ref{ia}) and (\ref{ib}) give useful cancellations in lines 1 and
3 of (\ref{lint3}). All but two
of the remaining differential operators (namely, the acceleration
terms) are
of the form (\ref{ic}) for suitable choice of $\alvec$ and $\betvec$.
Hence, we may repeatedly apply (\ref{ic}), yielding
\beq
\Li=\frac{q^2}{2\pi}\left[-\half(|\yd|^2+|\zd|^2)+(\dot{y}^\prl)^2+
(\dot{z}^\prl)^2+\dot{y}^\perp\dot{z}^\perp-\dot{y}^\prl\dot{z}^\prl\right]
\Lambda(|\yv-\zv|)+L_{\rm acc}
\label{lint5}
\eeq
where $L_{\rm acc}$ represents the remaining acceleration terms. In fact
\bea
L_{\rm acc}&=&-\frac{q^2}{2\pi}(\zdd\cdot\nabla_y+\ydd\cdot\nabla_z)
K_0(|\yv-\zv|)
=\frac{q^2}{2\pi}(\ydd-\zdd)\cdot(\yv-\zv)K_0'(|\yv-\zv|)\nonumber \\
&=&-\frac{q^2}{2\pi}\left[|\yd-\zd|^2\frac{K_0'(|\yv-\zv|)}{|\yv-\zv|}
+(\dot{y}^\prl-\dot{z}^\prl)^2\Lambda(|\yv-\zv|)\right]
+\begin{array}{c}\mbox{total time}\\ \mbox{derivative.}\end{array}
\label{lacc}
\eea
 Substituting
(\ref{lacc}) and (\ref{lam}) into (\ref{lint5}) finally yields
\beq
\label{lintf}
\Li=-\frac{q^2}{4\pi}|\yd-\zd|^2K_0(|\yv-\zv|) \,.
\eeq

Our point particle model of 2-vortex dynamics is completed by adding to
$\Li$ the usual nonrelativistic free Lagrangian for two particles of mass
$\pi$ (the vortex rest energy), so
\beq
L=\frac{\pi}{2}(|\yd|^2+|\zd|^2)-\frac{q^2}{4\pi}|\yd-\zd|^2K_0(|\yv-\zv|) \,.
\eeq
We may define centre of mass and relative coordinates
\beq
\Rv=\half(\yv+\zv),\qquad \rv=\half(\yv-\zv)
\eeq
so that
\beq
L=\pi|\dot{\Rv}|^2+\pi\left(1-\frac{q^2}{\pi^2}K_0(2|{\bf r}|)\right)|\dot{\rv}|^2 \,.
\eeq
Motion under this Lagrangian coincides with geodesic flow on the 2-vortex
moduli space with respect to the asymptotic metric (\ref{asymet}) of
section \ref{sec:svfm}: we simply identify $Z=R^1+iR^2$ and
$\sigma e^{i\theta}=r^1+ir^2$.

Extension of this treatment to the case of $N$ well separated vortices is
trivial due to the underlying linearity of our point-particle model. To the
standard Lagrangian for $N$ free particles of mass $\pi$, moving along
trajectories $\yv_r(t)$, $r=1,2,\ldots,N$ say,
one simply adds one copy of $\Li$, as in
eq.\ (\ref{lintf}), for each unordered pair of distinct particles, or
equivalently, one copy of $\half\Li$ for each ordered distinct pair. The
result is
\beq
\label{LN}
L=\frac{\pi}{2}\sum_r|\yd_r|^2-\frac{q^2}{8\pi}\sum_{r\neq s}
K_0(|\yv_r-\yv_s|)|\yd_r-\yd_s|^2 \,.
\eeq
The asymptotic formula for the $N$-vortex metric readily follows. Defining
complex coordinates $Z_r=y_r^1+iy_r^2$, their differences
$Z_{rs}=Z_r - Z_s$, and holomorphic 1-forms $dZ_{rs}=dZ_r-dZ_s$,
one sees that
\beq
\gmet =\pi\sum_r dZ_r d\bar Z_r-\frac{q^2}{4\pi}\sum_{r\neq
s}K_0(|Z_{rs}|) \, dZ_{rs} d\bar Z_{rs} \,,
\eeq
which coincides with (\ref{gN2}).

\section{Two-Vortex Scattering}
\setcounter{equation}{0}
\label{sec:tvs}

The relative motion of two vortices, in the geodesic approximation, is
determined by the purely kinetic Lagrangian
\be
L = \half \eta(\sigma) (\dot\sigma^2 + \sigma^2 \dot\theta^2)
\ee
where the ranges of $\sigma$ and $\theta$ are as in section 2. Samols
calculated the function $b(\sigma)$ and hence $\eta(\sigma)$
numerically, and using this found the
geodesic motion of two vortices \cite{Sa}. The geodesic motion has
two constants of integration: the energy $E$,
which is $L$ itself, and the angular momentum $\ell$, which equals
$\eta(\sigma)\sigma^2 \dot\theta$. Using these, one can find
$d\theta/d\sigma$, and the scattering angle can be
determined by integration. It depends on the impact parameter $a$,
given by the motion of
the vortices as they approach from infinity. There the energy is $\pi v^2$ and
the angular momentum is $2\pi av$.

The result for the scattering angle as a function of impact parameter
agrees well with numerical simulations of 2-vortex
scattering using the complete field equations \cite{KMR, SR}.
The motion is repulsive, and the
scattering angle increases monotonically as the impact parameter
decreases, from zero when the impact
parameter is infinite, up to $\frac{\pi}{2}$ in a head-on collision. The field
dynamics begins to significantly differ from the geodesic motion only
if the vortex speeds exceed about half the speed of light.

Now that we have obtained the asymptotic form of the 2-vortex metric,
with $\eta(\sigma)$ given by eq. (\ref{eta}), it is possible to estimate
the asymptotic scattering. Unfortunately, the exact relation between
scattering angle and impact parameter for this metric is given by a
rather intractable integral. Rather than
evaluate this numerically, we adopt a different strategy.

Since our asymptotic metric is only valid at large vortex separations,
it can only be used with confidence to evaluate the scattering angle
at large impact parameter. Here the vortex trajectories are almost
straight, and the scattering angle small. We can find the
approximate scattering angle by a perturbative calculation. It is
convenient to use Cartesian coordinates
$x= \sigma\cos\theta$ and $y=  \sigma\sin\theta$. The Lagrangian becomes
\be
L = \half \eta(\sigma) (\dot x^2 + \dot y^2)
\ee
where $\sigma = \sqrt{x^2 + y^2}$. The equations of motion are
\bea
\eta(\sigma) \ddot x + \frac{\eta'(\sigma)}{2\sigma}
(x \dot x^2 + 2y \dot y \dot x - x \dot y^2) &=& 0 \label{xdotdot} \\
\eta(\sigma) \ddot y + \frac{\eta'(\sigma)}{2\sigma}
(y \dot y^2 + 2x \dot x \dot y - y \dot x^2) &=& 0 \label{ydotdot} \,.
\eea

We may assume that the motion is approximately along the line $x=a$,
with $y$ increasing from $-\infty$ to $\infty$ at approximately
constant speed $v$. (This is in fact the trajectory of one of the
vortices; the other moves in the opposite direction along the line
$x=-a$.) The initial value of $\dot x$ is taken to be zero, and by
calculating the small change in $\dot x$ we can calculate the
scattering angle. We work to leading order in the small quantity
$\exp(-2a)$. $\eta'$ is of first order in $\exp(-2a)$ along the
trajectory, so at this order $\ddot x$ is given by the term
proportional to $\dot y^2$ in eq. (\ref{xdotdot}), and the coefficient
$\eta$ multiplying $\ddot x$ can be approximated by $2\pi$. $\eta'
\dot x$ is negligible. $\ddot y$ is of
order $\exp(-2a)$ too, but the consequent change of speed in the
$y$-direction along the trajectory can be neglected. Thus, it is a
sufficiently good approximation to take the solution of eq. (\ref{ydotdot})
to be $y=vt$, and to simplify eq. (\ref{xdotdot}) to
\be
\ddot x = \frac{av^2}{4\pi} \frac{\eta'(\sigma)}{\sigma} \,.
\ee
The total change in $\dot x$ is therefore
\be
\Delta \dot x = \frac{av^2}{4\pi} \int_{-\infty}^\infty
\frac{\eta'(\sigma)}{\sigma} \, dt \,,
\ee
and the scattering angle, assuming it is small, is
\be
\Theta = \frac{1}{v} \, \Delta \dot x \,.
\ee
Expressing $\Delta \dot x$ as an integral over $y$, using $y=vt$ and $\sigma
= \sqrt{a^2 + y^2}$, and also $\eta'(\sigma) = \frac{4q^2}{\pi}
K_1(2\sigma)$, we find
\bea
\Theta &=& \frac{q^2 a}{\pi^2} \int_{-\infty}^\infty
\frac{K_1(2\sqrt{a^2 + y^2})}{\sqrt{a^2 + y^2}} \, dy \\
&=& -\frac{q^2}{2\pi^2} \, \frac{d}{da} \int_{-\infty}^\infty
K_0(2\sqrt{a^2 + y^2}) \, dy \,.
\eea
Remarkably, this gives the simple result
\be
\Theta = \frac{q^2}{2\pi} \exp (-2a) \,.
\ee

The integral can be understood as follows. The planar Helmholtz
equation with source at the origin
\be
(-\nabla^2 + 4) \chi = 2\pi \delta(\xv)
\ee
has the exponentially decaying solution $\chi = K_0(2\sqrt{x^2 + y^2})$.
Integrating the equation with respect to $y$ we find that
$\tilde \chi = \int_{-\infty}^\infty K_0(2\sqrt{x^2 + y^2}) \, dy $
satisfies
\be
\left(-\frac{d^2}{dx^2} + 4 \right) \tilde\chi = 2\pi \delta(x)
\ee
and hence $\int_{-\infty}^\infty K_0(2\sqrt{x^2 + y^2}) \, dy
= \frac{\pi}{2} \exp(-2|x|)$.

\begin{figure}[t]
\centerline{\epsfysize=3truein
\epsfbox[100   220   520   570]{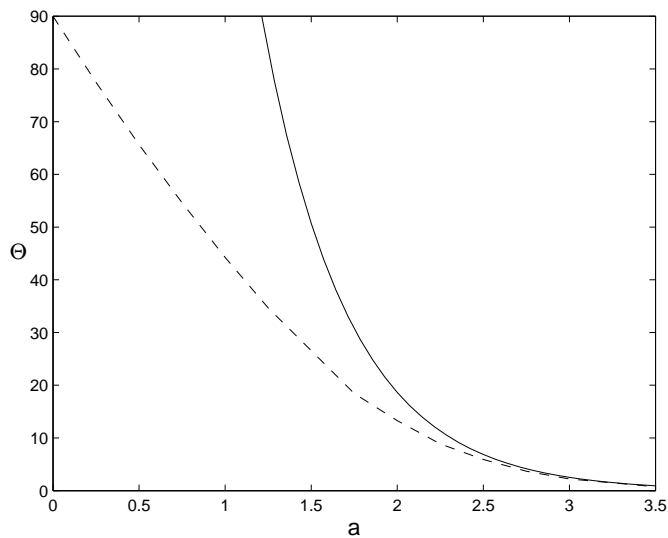}}
\caption{Scattering angle $\Theta$ against impact parameter $a$ for 2-vortex 
scattering in the geodesic approximation. Dashed curve: Samols' numerical
implementation; solid curve: our perturbative approximation.}
\label{figure}
\end{figure}

If we substitute the numerical value of $q$, we can compare the
dependence of scattering angle on impact parameter with Samols'
result. The agreement is good for $a \ge 2$. This is shown in Fig.\
\ref{figure}.

We expect corrections to this calculation. These are partly due to
the neglected terms in (\ref{xdotdot}) and (\ref{ydotdot}), and partly due
to corrections to our asymptotic metric.

\vskip 20pt
{\centerline{\bf{Acknowledgement}}}
\vspace{.25cm}
\noindent
We are grateful to David Tong for discussions and correspondence.

\end{document}